# Static forms of the Robertson-Walker spacetimes

**M Ibison**
Institute for Advanced Studies at Austin,
11855 Research Boulevard, Austin TX 78759-2443, USA

E-mail: ibison@earthtech.org

**Abstract.** It is shown that only the maximally-symmetric spacetimes can be expressed in both the Robertson-Walker form and in static form – there are no other static forms of the Robertson-Walker spacetimes. All possible static forms of the metric of the maximally-symmetric spacetimes are presented as a table. The findings are generalized to apply to functionally more general spacetimes: it is shown that the maximally symmetric spacetimes are also the only spacetimes that can be written in both orthogonal-time isotropic form and in static form.

## 1. Introduction

In the context of Cosmology, the fact that the de Sitter metric can be written in static form [1] permits novel re-interpretation of, for example, cosmological red-shift and 'expansion of spacetime', the more traditional descriptions of which are based on the time-evolving Robertson-Walker metrics. The aim of this effort is to determine to what extent this latitude in interpretation applies to other members of the Robertson-Walker family of spacetimes. It shares with [2] a common motivation, which is to help distinguish between the coordinate-dependent and coordinate-independent aspects of Cosmology.

Whilst the initial focus is to find all the static diagonal forms of the Robertson-Walker spacetimes, it turns out to be just as easy to pose a more general question whose answer reveals an interesting relationship between two quite general functional forms of the metric. One of these is described by any of the (obviously equivalent) line elements

$$\begin{aligned} ds^2 &= dt^2 - a^2(t)s^2(r)(dr^2 + r^2 d\Omega^2) &= a^2(t)(dt^2 - s^2(r)(dr^2 + r^2 d\Omega^2)) \\ &= dt^2 - a^2(t)(s^2(r)dr^2 + r^2 d\Omega^2) &= a^2(t)(dt^2 - s^2(r)dr^2 - r^2 d\Omega^2) \\ &= dt^2 - a^2(t)(dr^2 + s^2(r)d\Omega^2) &= a^2(t)(dt^2 - dr^2 - s^2(r)d\Omega^2) \end{aligned} \quad \begin{pmatrix} a,b \\ 1 \; c,d \\ e,f \end{pmatrix}$$

where $s$ and $a$ are arbitrary functions (not the same in each case). These obviously equivalent spacetimes will be denoted OTI standing for 'orthogonal time isotropic' and more formally by $\{\text{OTI}(t)(r)\}$ where $(t)$ and $(r)$ stand respectively for the functional degrees of freedom $a^2(t)$ and $s^2(r)$ in any of the obviously equivalent forms in (1).

The other family of metrical forms under consideration is that which can be written in any of the (obviously equivalent) diagonal forms

$$\begin{aligned}
\mathrm{d}s^2 &= A^2(R)\left(\mathrm{d}T^2 - \mathrm{d}R^2\right) - B^2(R)\mathrm{d}\Omega^2 \\
&= A^2(R)\mathrm{d}T^2 - B^2(R)\mathrm{d}R^2 - R^2\mathrm{d}\Omega^2 \\
&= A^2(R)\mathrm{d}T^2 - B^2(R)\left(\mathrm{d}R^2 + R^2\mathrm{d}\Omega^2\right) \\
&= A^2(R)\mathrm{d}T^2 - \mathrm{d}R^2 - B^2(R)\mathrm{d}\Omega^2
\end{aligned}$$

(2 a,b,c,d)

where $A$ and $B$ are arbitrary functions (not the same in each case). With reference specifically to the particular form (2c), these spacetimes will be denoted SI standing for 'static isotropic' and more formally by $\{\mathrm{SI}(r)(r)\}$ where each $(r)$ stands for a single functional degree of freedom. Throughout this paper it will be understood that the coordinates and possibly the interval $\mathrm{d}s$ have been normalized with respect to a fixed distance; the possibility of a real linear transformation of the coordinates $x \to x' = ax + b$, $a,b$ real will be understood.

The Robertson-Walker metrics are a subset of (1) with, say

$$s(r) = \sin\left(\sqrt{K}r\right)/\sqrt{K} \equiv S_K(r); \quad K = [1,0,-1] \tag{3}$$

in (1c). The collection of Robertson-Walker spacetimes will be denoted by $\{RW_K(t)\}$, which includes all three cases $K = [-1,0,1]$ and where $(t)$ stands for the functional degree of freedom that in this case is the RW scale factor $a(t)$. Spacetimes that are equivalent to a particular RW spatial geometry will be identified with a subscript so that, for example, the hyperbolic space form of the RW metric generates spacetimes $\{RW_{-1}(t)\}$. Therefore, allowing for some overlap between the subsets [CQG conformal],

$$\{RW_K(t)\} = \{RW_{-1}(t)\} \cup \{RW_0(t)\} \cup \{RW_{+1}(t)\}. \tag{4}$$

In this document a 'spacetime' means a pseudo-Riemannian manifold. No distinction is made between different topologies that are otherwise locally equivalent through a coordinate transformation. All metrics under consideration here are diagonal, and are most conveniently expressed as a line element. Discussion of the extensibility of a geometry and its covering of a given spacetime is avoided.

## 2. Static form of Robertson-Walker spacetimes

*2.1 Allowed forms for the transformation*

In making the correspondence between $\{\mathrm{OTI}(t)(r)\}$ and $\{\mathrm{SI}(r)(r)\}$ initially it will be easier to work specifically with (1d) and (2b). Then the relationship between the new and old coordinates is

$$\mathrm{d}T = T_t \mathrm{d}t + T_r \mathrm{d}r, \quad \mathrm{d}R = R_t \mathrm{d}t + R_r \mathrm{d}r. \tag{5}$$

Inserting this into (2b) and equating with (1d) gives

$$\begin{aligned}
A^2 T_t^2 - B^2 R_t^2 &= a^2 \\
B^2 R_r^2 - A^2 T_r^2 &= a^2 s^2 \\
A^2 T_t T_r - B^2 R_t R_r &= 0
\end{aligned} \tag{6}$$

and

$$R^2 = a^2 r^2 \Rightarrow R = a(t)r. \tag{7}$$

Putting (7) into the other three gives

$$A^2 T_t^2 - B^2 \dot{a}^2 r^2 = a^2$$
$$B^2 a^2 - A^2 T_r^2 = a^2 s^2 \ . \tag{8}$$
$$A^2 T_t T_r - B^2 \dot{a} a r = 0$$

Solving for $A$ from the last of (8):

$$A^2 = \frac{B^2 \dot{a} a r}{T_t T_r} \ . \tag{9}$$

Inserting in the other two equations gives

$$B^2 \dot{a} a r \xi - B^2 \dot{a}^2 r^2 = a^2; \quad \xi \equiv T_t / T_r$$
$$B^2 a^2 - B^2 \dot{a} a r / \xi = a^2 s^2 \tag{10}$$

Together these give

$$s^2 = \frac{a^2 - \dot{a} a r / \xi}{\dot{a} a r \xi - \dot{a}^2 r^2} = \frac{a}{\dot{a} r \xi} \Rightarrow \frac{T_r}{T_t} = \frac{\dot{a}(t)}{a(t)} r s^2(r) . \tag{11}$$

The general solution of (11) can be written down immediately:

$$T = F\left( \int^r dr' r' s^2(r') + \int^t dt' a(t') / \dot{a}(t') \right) \tag{12}$$

where $F$ is an arbitrary function. From (10) the equation for $B$ is then

$$B^2(R) = B^2(ar) = \frac{a^2}{\dot{a} a r \xi - \dot{a}^2 r^2} = \frac{1}{1/s^2 - (\dot{a} r / a)^2} \ . \tag{13}$$

Eqs. (7) and (12) determine the transformation relating (1d) to (2b). However, eq. (13) turns out to be sufficient to determine (up to a few constants) the functions $A(R), B(R), s(r)$ and $a(t)$, as is shown in the following.

## 2.2 Compatible subsets of $\{SI(r)(r)\}$ and $\{OTI(r)(r)\}$

Here we determine the constraints on the functional degrees of freedom in $\{SI(r)(r)\}$ and $\{OTI(r)(r)\}$ in order that the resulting subsets are equivalent and therefore describe the same spacetime.

The solution of eq. (13) is aided by re-writing it as

$$f(ar) = g(r) - r^2 h(a) \tag{14}$$

where

$$f \equiv 1/B^2, \quad g \equiv 1/s^2, \quad h \equiv (\dot{a}/a)^2 . \tag{15}$$

The unknown function $f$ can be eliminated by differentiation to give

$$r \frac{\partial f}{\partial r} - a \frac{\partial f}{\partial a} = 0 = rg' + 2r^2 h - r^2 a h' . \tag{16}$$

The right hand side is separable into the sum of two independent one-parameter equations:

$$g'(r)/r = ah'(a) - 2h(a) = -2\alpha \tag{17}$$

where $\alpha$ is a constant. The solutions are

$$g(r) = \beta - \alpha r^2, \quad h(a) = \gamma a^2(t) - \alpha \tag{18}$$

and therefore

$$f = \beta - \gamma(ar)^2. \tag{19}$$

Restoring from (14) the original functions $B$ and $s$, (18) and (19) give

$$B^2(R) = \frac{1}{\beta - \gamma R^2}, \quad s^2(r) = \frac{1}{\beta - \alpha r^2}. \tag{20}$$

And the equations for $h$ in (15) and (18) give

$$(\dot{a}/a)^2 = \gamma a^2(t) - \alpha. \tag{21}$$

The solution to this equation is easier to recognize on letting $\phi = 1/a$

$$\left(\frac{d\phi}{dt}\right)^2 + \alpha \phi^2 = \gamma \tag{22}$$

whereupon it is clear that

$$\phi = \sqrt{\frac{\gamma}{\alpha}} \sin(\sqrt{\alpha} t + \chi) \Rightarrow a(t) = \frac{\sqrt{\alpha}}{\sqrt{\gamma} \sin(\sqrt{\alpha} t + \chi)} \tag{23}$$

where $\chi$ is a constant. Note that the constraint that $a(t)$ is real still permits $\alpha < 0$.

We now compute the argument of $F$ in (12). Using the results (20) and (23) for $s(r)$ and $a(t)$ respectively one finds

$$\int^r dr' r' s^2(r') = \int^r dr' \frac{1}{\beta - \alpha r'^2} = -\frac{1}{2\alpha} \log(\beta - \alpha r^2), \tag{24}$$

and

$$\int^t dt' \dot{a}(t')/a(t') = -\int^t dt' \frac{1}{\sqrt{\alpha}} \tan(\sqrt{\alpha} t + \chi) = \frac{1}{\alpha} \log \cos(\sqrt{\alpha} t + \chi). \tag{25}$$

Combining these into (12) gives

$$T = G\left(\frac{\cos(\sqrt{\alpha} t + \chi)}{\sqrt{\beta - \alpha r^2}}\right) \tag{26}$$

where $G$ is a new arbitrary function.

Eq. (9) can now be used to furnish a functional equation sufficient to determine both of the functions $A$ and $G$. From (26) we have

$$T_r = \frac{\alpha r}{\left(\beta - \alpha r^2\right)^{3/2}} \cos\left(\sqrt{\alpha}t + \chi\right)G', \quad T_t = -\sqrt{\alpha} \frac{\sin\left(\sqrt{\alpha}t + \chi\right)}{\sqrt{\beta - \alpha r^2}} G'$$

$$\Rightarrow T_r T_t = -\alpha^{3/2} \frac{r \sin\left(\sqrt{\alpha}t + \chi\right) \cos\left(\sqrt{\alpha}t + \chi\right)}{\left(\beta - \alpha r^2\right)^2} G'^2 \tag{27}$$

Using this and obtaining $B$ from (20) and $a$ from (23), (9) now gives

$$A^2(R) = \frac{\left(\beta - \alpha r^2\right)^2}{\gamma\left(\beta - \gamma R^2\right)\sin^4\left(\sqrt{\alpha}t + \chi\right)G'^2} . \tag{28}$$

Since the left hand side is a function of $R$ only, a method of solution is to write the right hand side in terms of $R$ and one other independent variable, and then require that the functional dependence on the latter vanish. (It is not necessary to choose $T$ for the other variable, it is only necessary that it does not depend only on $R$.) Here we choose $a(t)$ as the other variable, and rewrite the harmonic functions accordingly:

$$A^2(R) = \frac{\gamma\left(\beta - \alpha r^2\right)^2 a^4}{\alpha^2\left(\beta - \gamma R^2\right)} \frac{1}{G'^2} = \frac{\left(a^2\beta/\alpha - R^2\right)^2}{\beta/\gamma - R^2} \frac{1}{G'^2} . \tag{29}$$

It remains to express $G$ in terms of $R$ and $a$. It is convenient to let

$$1/G'^2(z) = H(z^2) \tag{30}$$

where $H$ is a function to be determined, and

$$z^2 \equiv \frac{\cos^2\left(\sqrt{\alpha}t + \chi\right)}{\beta - \alpha r^2} = \frac{1 - \alpha/\gamma a^2}{\beta - \alpha R^2/a^2} = \frac{1}{\beta}\left(\frac{a^2\beta/\alpha - \beta/\gamma}{a^2\beta/\alpha - R^2}\right). \tag{31}$$

Putting this into (28) gives

$$A^2(R) = \frac{\left(a^2\beta/\alpha - R^2\right)^2}{\beta/\gamma - R^2} H\left(\frac{1}{\beta}\left(\frac{a^2\beta/\alpha - \beta/\gamma}{a^2\beta/\alpha - R^2}\right)\right). \tag{32}$$

The problem now is to find functions $A$ and $H$ such that this equation is true for any $R$ and $a$. Let

$$x = R^2, \quad y = a^2\beta/\alpha, \quad f(x) = \sqrt{\mu - x}A\left(\sqrt{x}\right), \quad h(z) = \sqrt{H(z/\beta)}, \quad \mu = \beta/\gamma. \tag{33}$$

Then (32) becomes the functional equation

$$f(x) = (y - x)h\left(\frac{y - \mu}{y - x}\right) \quad \forall (x, y) \tag{34}$$

($\mu$ held fixed). This equation can further be reduced with a few simple substitutions:

$$x \to x+\mu, \; y \to y+\mu \qquad \Rightarrow \quad f(x+k) = (y-x)h\left(\frac{y}{y-x}\right)$$

$$y \to y+x \qquad \Rightarrow \quad f(x+k) = yh(1+x/y) \qquad (35)$$

$$z = x/y, \; f_1(x) = f(x+\mu), \; h_1(x) = h(1+x) \quad \Rightarrow \quad f_1(yz) = yh_1(z)$$

The general solution to the last equation is easily written down:

$$f_1(x) = -\lambda x, \quad h_1(x) = -\lambda x \qquad (36)$$

where $\lambda$ is an arbitrary constant. The original functions $f$ and $g$ in (34) are found by reversing the substitutions made in (35):

$$f(x) = \lambda(\mu - x), \quad h(x) = \lambda(1-x). \qquad (37)$$

Similarly, the function $A$ is found by reversing the substitutions in (33):

$$A(R) = \frac{f(R^2)}{\sqrt{\mu - R^2}} = \frac{\lambda(\mu - R^2)}{\sqrt{\mu - R^2}} = \lambda\sqrt{\beta/\gamma - R^2}. \qquad (38)$$

The function $H$ is found by reversing the substitutions in (33) linking $h$ to $H$

$$H(z) = h^2(\beta z) = \lambda^2(1-\beta z)^2. \qquad (39)$$

Recalling (30), this gives

$$G'(z) = \frac{1}{\sqrt{H(z^2)}} = \frac{1}{\lambda(1-\beta z^2)} \Rightarrow G(z) = \frac{1}{\lambda}\sqrt{\frac{\alpha}{\beta}}\tanh^{-1}(\sqrt{\beta}z) + G_0. \qquad (40)$$

Finally, (26) gives

$$T = \frac{1}{\lambda\sqrt{\beta}}\tanh^{-1}\left(\frac{\cos(\sqrt{\alpha}t + \chi)}{\sqrt{1-r^2\alpha/\beta}}\right) + G_0 \qquad (41)$$

*2.3 Identification of the maximally-symmetric spacetimes*

In summary, the demand that a spacetime be expressible in both $\{SI(r)(r)\}$ and $\{OTI(r)(r)\}$ fixes the functional degrees of freedom $A(R), B(R), s(r)$ and $a(t)$ to that given respectively by (38), (20), and (23), so that (1d) must have the form

$$ds^2 = \frac{\alpha}{\gamma \sin^2(\sqrt{\alpha}t + \chi)}\left(dt^2 - \frac{dr^2}{\beta(1-r^2\alpha/\beta)} - r^2 d\Omega^2\right) \qquad (42)$$

and (2b) must have the form

$$ds^2 = \lambda^2(\beta/\gamma - R^2)dT^2 - \frac{dR^2}{\beta(1-R^2\gamma/\beta)} - R^2 d\Omega^2, \qquad (43)$$

wherein the relationship between the coordinates is given by

$$R = r \frac{\sqrt{\alpha}}{\sqrt{\gamma} \sin(\sqrt{\alpha} t + \chi)}, \quad T = \frac{1}{\lambda \sqrt{\beta}} \tanh^{-1}\left(\frac{\cos(\sqrt{\alpha} t + \chi)}{\sqrt{1 - r^2 \alpha / \beta}}\right) + G_0 \quad (44)$$

where all the lower case Greek symbols are constants.

The constants are constrained by the requirement that the transformation is real and does not change the signature. From Eq. (43) it is clear (at $R = 0$) that $\beta > 0$, and consequently that $\mathrm{sgn}(\gamma) = \mathrm{sgn}(\lambda^2)$. Bearing this in mind the coordinate changes

$$\mathrm{d}\Omega^2 \to \mathrm{d}\Omega^2 / |\beta|, \quad \mathrm{d}T^2 \to \mathrm{d}T^2 / |\lambda^2||\beta|, \quad R^2 \to R^2 |\beta| / |\gamma|, \quad K = \mathrm{sgn}(\gamma) \quad (45)$$

convert the line element (43) to

$$|\gamma| \mathrm{d}s^2 = (1 - KR^2) \mathrm{d}T^2 - \frac{\mathrm{d}R^2}{1 - KR^2} - R^2 \mathrm{d}\Omega^2 \quad (46)$$

where $K = \pm 1$, which are recognized to be the static form of the maximally symmetric spacetimes [1] (Tolman refers to the anti-de Sitter universe as a closed de Sitter universe). The corresponding Robertson-Walker line-element in normalized coordinates is achieved with

$$\mathrm{d}\Omega^2 \to \mathrm{d}\Omega^2 / |\beta|, \quad \mathrm{d}t^2 \to \mathrm{d}t^2 / |\alpha|, \quad r^2 \to r^2 |\beta| / |\alpha|, \quad \kappa = \mathrm{sgn}(\alpha), \quad (47)$$

which converts the line element (42) to

$$|\gamma| \mathrm{d}s^2 = \frac{\kappa K}{\sin^2(\sqrt{\kappa} t + \chi)} \left( \mathrm{d}t^2 - \frac{\mathrm{d}r^2}{1 - \kappa r^2} - r^2 \mathrm{d}\Omega^2 \right) \quad (48)$$

where $\kappa = \pm 1$. The re-scalings (47) and (45) together with the substitution $\lambda = |\lambda| e^{i\psi}$ ($\psi$ real), changes the transformations to

$$R = r \frac{\sqrt{K} \sqrt{\kappa}}{\sin(\sqrt{\kappa} t + \chi)}, \quad T = e^{i\psi} \tanh^{-1}\left(\frac{\cos(\sqrt{\kappa} t + \chi)}{\sqrt{1 - \kappa r^2}}\right) + G_0. \quad (49)$$

It is necessary to consider as separate cases the possibilities $\kappa = \pm 1$ and $K = \pm 1$ that characterize the transformation (49) relating the Robertson-Walker form of the line-element (48) to the static form (46).

*2.4 Different cases*

$\underline{\kappa = 1, K = 1}$

Positive real $R$ is achieved if $r < 1$ and $0 < t + \chi < \pi$. Real $T$ then requires $\psi = 0$ or $\pi$. Let us choose $\chi = \pi / 2$, $\psi = \pi$, and $G_0 = 0$ so that, up to additional offsets in $t$ and $T$, (49) becomes

$$R = \frac{r}{\cos t}, \quad T = \tanh^{-1}\left(\frac{\sin t}{\sqrt{1 - r^2}}\right) \quad (50)$$

which is valid for $-\pi/2 < t < \pi/2$. Then (46) is

$$\mathrm{d}s^2 = (1 - R^2) \mathrm{d}T^2 - \frac{\mathrm{d}R^2}{1 - R^2} - R^2 \mathrm{d}\Omega^2 \quad (51)$$

which is a line element of the de Sitter spacetime (the common factor of $|\gamma|$ is absorbed into a normalized $ds$). Correspondingly, (48) is

$$ds^2 = \frac{1}{\cos^2 t}\left(dt^2 - \frac{dr^2}{1-r^2} - r^2 d\Omega^2\right). \tag{52}$$

Other static forms of the line element (51) corresponding to the different possibilities in (2) are obtained by a coordinate transformation of the radial coordinate alone. All three possibilities are given in table 1.

<u>$\kappa = 1, K = -1$</u>

This case has no solutions for the available constants which leave $R$ real for a range of $t$.

<u>$\kappa = -1, K = 1$</u>

For real $R$ this combination requires $\chi = i|\chi|$ where the offset $|\chi|$ can then be absorbed into the definition of $t$. This gives

$$R = \frac{r}{\sinh t}, \quad T = e^{i\psi}\tanh^{-1}\left(\frac{\cosh t}{\sqrt{1+r^2}}\right) + G_0. \tag{53}$$

The corresponding line element in the new coordinates is unchanged from (51). Therefore this is the de Sitter spacetime, entries for which are given in table 1. Eq. (51) is valid only for $R \in [0,1]$ and the relationship (53) between $R$ and $r$ determines that $\cosh t \geq \sqrt{1+r^2}$, which causes the $\tanh^{-1}$ function in (53) to become complex. In order to ensure that $T$ is real, we use the result that for any integer $n$

$$\tanh^{-1} x = \tanh^{-1} 1/x + (n+1/2)i\pi, \tag{54}$$

so that for suitable choice of $\psi$ and $G_0$ (53) becomes

$$R = \frac{r}{\sinh t}, \quad T = \tanh^{-1}\left(\frac{\sqrt{1+r^2}}{\cosh t}\right). \tag{55}$$

The line element in the RW coordinates is

$$ds^2 = \frac{1}{\sinh^2 t}\left(dt^2 - \frac{dr^2}{1+r^2} - r^2 d\Omega^2\right) \tag{56}$$

(the factor of $|\gamma|$ has been absorbed into $ds$). This and obviously equivalent entries corresponding to other entries in (1) are given in table 1.

<u>$\kappa = -1, K = -1$</u>

For positive real $R$ this combination requires $\chi = \pi/2 + i\phi$ where then a real $\phi$ can be absorbed into the definition of $t$. Real $T$ is then achieved with $\psi = \pi/2$ whereupon

$$R = \frac{r}{\cosh t}, \quad T = \tan^{-1}\left(\frac{\sinh t}{\sqrt{1+r^2}}\right). \tag{57}$$

The line element (46) is then

$$ds^2 = (1+R^2)dT^2 - \frac{dR^2}{1+R^2} - R^2 d\Omega^2 \tag{58}$$

and the line element (48) is

$$ds^2 = \frac{1}{\cosh^2 t}\left(dt^2 - \frac{dr^2}{1+r^2} - r^2 d\Omega^2\right) \tag{59}$$

which is the anti-de Sitter spacetime. (The factor of $|\gamma|$ has been absorbed into $ds$.) Other static forms of the line element (58) corresponding to the different possibilities in (2) are obtained by a coordinate transformation of the radial coordinate alone. All three possibilities are given in table 1.

## 3. Discussion

*3.1 Relationship between spacetimes*
The only spacetimes that are simultaneously expressible in SI and OTI form are the maximally symmetric spacetimes. Therefore the latter can be uniquely identified by demanding that the line element be expressible in any of the forms in (1) and in any of the forms in (2). Symbolically:

$$\{\text{OTI}(t)(r)\} \cap \{\text{SI}(r)(r)\} = \{\text{de Sitter, Minkowski, anti-de Sitter}\}$$
$$\equiv \{\text{maximally symmetric}\} \tag{60}$$

Since the Robertson-Walker spacetimes are a subset of $\{\text{OTI}(t)(r)\}$, and contain the maximally symmetric spacetimes as special cases

$$\{\text{OTI}(t)(r)\} \supset \{\text{RW}_K(t)\} \supset \{\text{maximally symmetric}\}. \tag{61}$$

it obviously follows that the maximally symmetric spacetimes are those that can be expressed in static isotropic form and in Robertson-Walker form:

$$\{\text{RW}_K(t)\} \cap \{\text{SI}(r)(r)\} = \{\text{maximally symmetric}\}. \tag{62}$$

The relations (60), (61) and (62) is illustrated in figure 1.

*3.2 Coordinate-dependent interpretations of Cosmology*
    A product of this analysis is the catalogue of line elements in table 1 together with coordinate transformations linking the different forms in the body of this document, plus the determination of relationships between some general spacetimes, as given in figure 1.
    It is hoped that some of the utility of the effort is in highlighting the coordinate dependence of traditional interpretations of Cosmology. The static isotropic version of the de Sitter spacetime fits the same observational data that, interpreted in the traditional Robertson-Walker coordinates with line-element (1a) say, gives rise to the exponentially increasing scale factor – to which our future is (presently) considered to be asymptotic. For the duration of this exponential era one may therefore apply the static coordinate system with line element (51) (i.e. as a patch), leading to an unconventional, but equally applicable, interpretation of Cosmological redshift. Notice first that the line element (51) is the same as the Schwarzschild line element with the Newtonian potential $\phi(r) = -2GM/r$ replaced by $\phi(r) = -r^2/\Lambda^2$ where $\Lambda$ is some constant length to be determined from the observational data. This is a quadratic repulsive potential which, relatively locally (when $r^2/\Lambda^2 \ll 1$) characterizes a constant repulsive force in the Newtonian theory. In this coordinate system Cosmological red-shift is not,

therefore, a $a^2(t)$ consequence of expansion of space, it is effectively an inverse gravitational red-shift wherein incoming light must climb a potential hill in order to reach the observer.

**Table 1.** Catalogue of diagonal line elements discussed in this document.

$$\{\text{OTI}(t)(r)\}$$

| line element | | text references | |
|---|---|---|---|
| $dt^2 - a^2(t)s^2(r)(dr^2 + r^2 d\Omega^2)$ | | (1a) | |
| $a^2(t)(dt^2 - s^2(r)(dr^2 + r^2 d\Omega^2))$ | | (1b) | |
| $dt^2 - a^2(t)(dr^2 + s^2(r) d\Omega^2)$ | | (1c) | |
| $a^2(t)(dt^2 - dr^2 - s^2(r) d\Omega^2)$ | | (1d) | |
| $dt^2 - a^2(t)(dr^2 + s^2(r) d\Omega^2)$ | | (1e) | |
| $a^2(t)(dt^2 - dr^2 - s^2(r) d\Omega^2)$ | | (1f) | |

$$\{\text{SI}(r)(r)\}$$

| line element | | text references | |
|---|---|---|---|
| $a^2(r)(dt^2 - dr^2) - b^2(r) d\Omega^2$ | | (2a) | |
| $a^2(r)dt^2 - b^2(r)dr^2 - r^2 d\Omega^2$ | | (2b) | |
| $a^2(r)dt^2 - b^2(r)(dr^2 + r^2 d\Omega^2)$ | | (2c) | |
| $a^2(r)dt^2 - dr^2 - b^2(r) d\Omega^2$ | | (2d) | |

$$\{\text{anti-de Sitter}\}$$

| line element | coordinate system | text references | literature references |
|---|---|---|---|
| $(1+r^2)dt^2 - \dfrac{dr^2}{1+r^2} - r^2 d\Omega^2$ | static | (58) | [1] |
| $\left(\dfrac{1+\mathbf{x}^2}{1-\mathbf{x}^2}\right)^2 dt^2 - \dfrac{d\mathbf{x}^2}{(1-\mathbf{x}^2)^2}$ | static isotropic | from (58) | |
| $\dfrac{1}{\cos^2 r}(dt^2 - dr^2 - \sin^2 r \, d\Omega^2)$ | static | from (58) | [3,4][1] |
| $\cosh^2 r \, dt^2 - dr^2 - \sinh^2 r \, d\Omega^2$ | static | from (58) | [3,4] |

---

[1] There is a mistake in [4].

| | line element | coordinate system | text references | literature references |
|---|---|---|---|---|
| | {de Sitter} | | | |
| | $\left(1-r^2\right)\mathrm{d}t^2 - \dfrac{\mathrm{d}r^2}{1-r^2} - r^2\mathrm{d}\Omega^2$ | static | (51) | [1,5] |
| | $\left(\dfrac{1-\mathbf{x}^2}{1+\mathbf{x}^2}\right)^2 \mathrm{d}t^2 - \dfrac{\mathrm{d}\mathbf{x}^2}{\left(1+\mathbf{x}^2\right)^2}$ | static isotropic | from (51) | |
| | $\dfrac{1}{\cosh^2 r}\left(\mathrm{d}t^2 - \mathrm{d}r^2 - \sinh^2 r\, \mathrm{d}\Omega^2\right)$ | static | from (51) | |
| | $\cos^2 r\, \mathrm{d}t^2 - \mathrm{d}r^2 - \sin^2 r\, \mathrm{d}\Omega^2$ | static | from (51) | |

**Notes on the table entries**

- It is understood that every line element can be scaled by an arbitrary constant (conformal) scale factor without changing the classification of the spacetime. Further, all coordinates are understood to be normalized with respect to some fixed length (which is in principle observable). That is, for a general diagonal line element

$$\mathrm{d}s^2 = \sum_i g_{ii}\left(x^0, x^1, x^2, x^3\right)\left(\mathrm{d}x^i\right)^2.$$

The classification is unchanged if

$$\mathrm{d}s^2 \to \alpha \sum_i g_{ii}\left(\beta_0 x^0, \beta_1 x^1, \beta_2 x^2, \beta_3 x^3\right)\left(\beta_i \mathrm{d}x^i\right)^2.$$

(The arbitrary rescaling of the interval $\mathrm{d}s$ is not generally the same as a coordinate transformation.) It follows that the trigonometric functions sine and cosine can be interchanged.

- Everywhere $a(t), s(r), a(r), b(r)$ are arbitrary functions, each occurrence of which is to be treated de novo.

**Figure 1.** The set of three maximally-symmetric spacetimes is the intersection between the set of orthogonal time isotropic spacetimes and the set of static isotropic spacetimes. The Schwarzschild spacetime is placed in context.

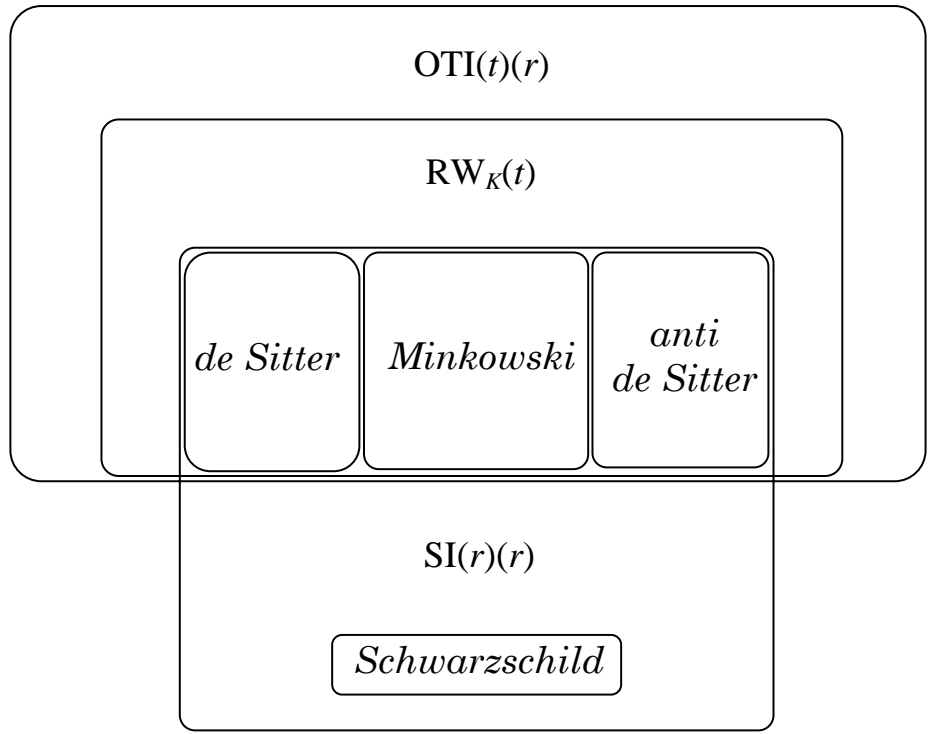